\documentclass{iau} 
\usepackage{graphicx}

\title[JD 1.~~Helium variations in GCs] %% give here short title %%
{Helium variations in Galactic and \\ extragalactic Globular Clusters}

\author[Edoardo P. Lagioia et al.]   %% give here short author list %%
{Edoardo P. Lagioia$^1$, Antonino P. Milone$^1$, Anna F. Marino$^1$, Aaron Dotter$^2$, Giacomo Cordoni$^1$ \and Marco Tailo$^1$}

\affiliation{$^1$Department of Physics and Astronomy, University of Padua, 35121 \\ 
Via Marzolo 8, Padua, Italy \\ email: {\tt edoardo.lagioia@unipd.it} \\[\affilskip]
$^2$Center For Astrophysics, Harvard \& Smithsonian, 60 Garden Street, 02138, \\
Cambridge, MA, USA}

\pubyear{2019}
\volume{351}  
\jname{Star Clusters: From the Milky Way to the Early Universe}
\editors{A. Bragaglia, M.B. Davies, A. Sills \& E. Vesperini, eds.}
\begin{document}

\maketitle
%. CONTINUE EDITING FROM HERE

\begin{abstract}
	The recent measurements of internal variations of helium in Galactic
	and extragalactic Globular Clusters (GCs) set binding constraints to
	the models of formation of Multiple Populations (MPs) in GCs, and gave
	rise, at the same time, to crucial questions related with the influence
	of the environment on MP formation as well as with the role played by
	GCs in the early galactic formation. We present the most recent
	estimates of helium enrichment in the main populations of a large
	sample of Galactic and extragalactic GCs. %, discussing their
	%implications for the theoretical predictions about the origin of MPs.
\end{abstract}

\firstsection % if your document starts with a section,
              % remove some space above using this command.I don
\section{Introduction}
Multiple stellar populations (MPs) in old Galactic Globular Clusters (GCs) are
characterized by a different content of light-elements, with well defined
abundance patterns like C-N and Na-O anti-correlations, by different helium
mass content ($\delta$Y) and, in few cases, also by different metallicity (for
reviews see \cite[Gratton et al. 2012]{gratton12}; \cite[Marino et al.
2019]{marino19}). 

The determination of internal helium variations in GCs is of primary importance
because it provides compelling information on the processes that led to the
formation of MPs and, ultimately, gives important clues on the role of the
first stars in the reionization of the Universe (see \cite[Lagioia et al.
2018]{lagioia18}; \cite[Milone et al.  2018]{milone18}). 

Recently, the discovery of MP in extragalactic GCs opened up new paths of
investigation for the chemical anomalies in clusters with different ages and
physical properties. 

We exploited multi-wavelength high-resolution \textit{Hubble Space Telescope}
(\textit{HST}) observations to perform, for the first time, an homogeneous
measurement of $\delta$Y between MPs in a large sample of Galactic GCs, using
the luminosity of Red Giant Branch (RGB) bump stars. Moreover, for the first
time, we measured light-elements and helium variations in MPs of four
extragalactic GCs belonging to the Small Magellanic Cloud (SMC). 

We found that Galactic and extragalactic GCs share similar properties, with
second generation (2G) stars enhanced in helium and nitrogen, and depleted in
carbon and oxygen, with respect to the pristine stellar generation (1G).

%\section{First and second generation stars in Globular Clusters}
\section{The RGB bump of multiple stellar populations}
The Ultraviolet (UV) Legacy Survey of Galactic GCs observed 57 clusters in the
three broad-band filters F275W, F336W, and F438W, available at the UV and
Visual Channel of the Wide Field Camera 3 (WFC3/UVIS) onboard \textit{HST}
(\cite[Piotto et al. 2015]{piotto15}). Appropriate combinations of these three
bands are at the base of the construction of photometric diagrams suitable for
the identification of stellar populations characterized by a different content
of light elements (\cite[Milone et al. 2015]{milone15}).  The location of the
RGB stars in such diagrams indicates that every cluster has two main stellar
populations: the first or 1G, with halo-like content of carbon, nitrogen, and
oxygen; the second or 2G, depleted in C and O and enhanced in N with respect to
the 1G. 

Similar diagrams, but based on the combination of the UVIS/WFC3 narrow-band filter
F343N and broad-band filters F336W and F438W, as well as of the ACS/WFC
broad-band filters F555W and F814W, allowed us to identify the two main
populations in the four Small Magellanic Cloud (SMC) GCs, NGC\,121, NGC\,339,
NGC\,416 and Lindsay\,1 (\cite[Lagioia et al. 2019]{lagioia19}).

\begin{figure}
\centering
\includegraphics[width=.4\textwidth]{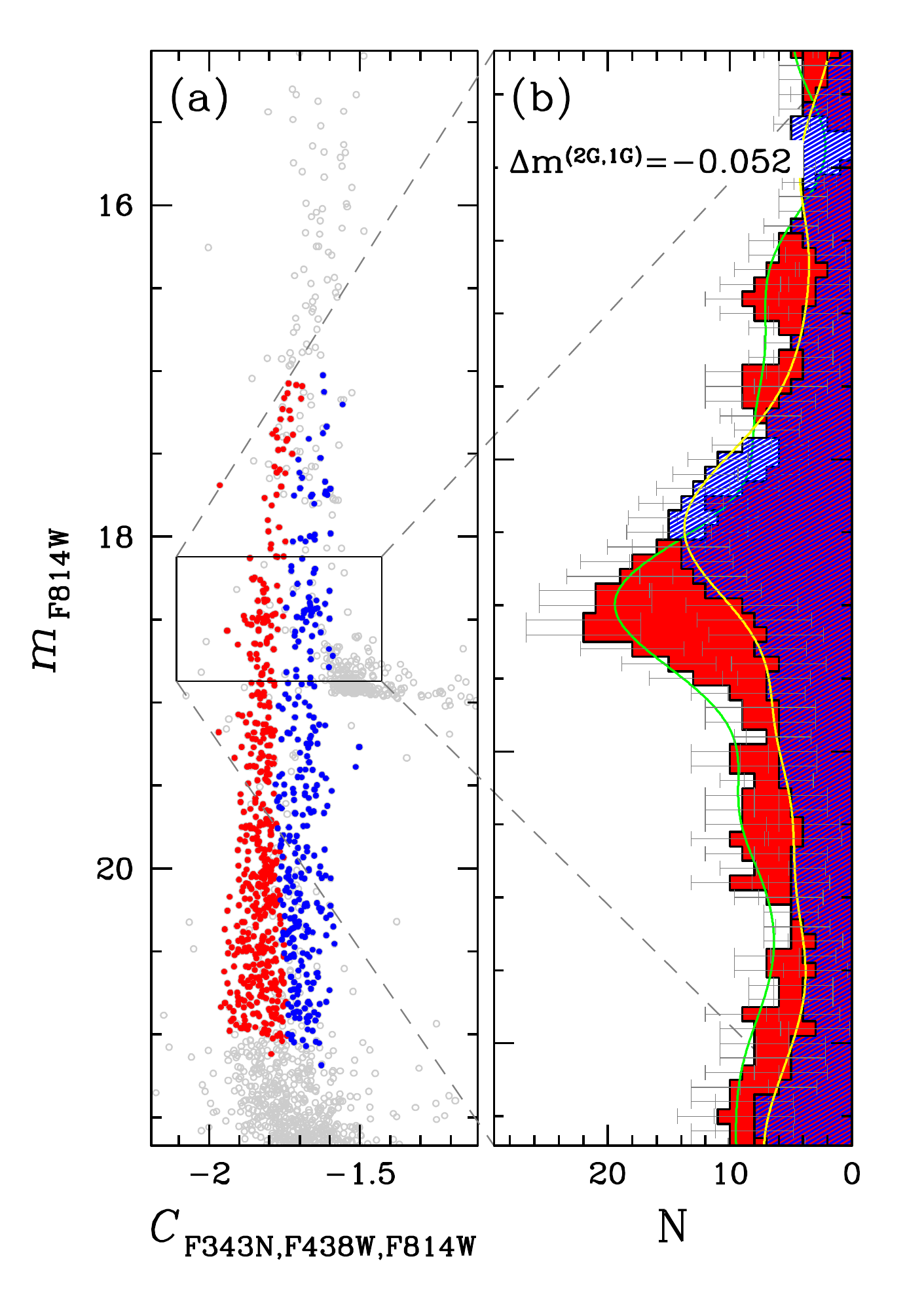}
\caption{\textit{Panel (a)}: $\mathrm{m_{F814W}}$ vs.
	$\mathrm{C_{F343N,F438W,F814W}}$ pseudo CMD of the 1G (red) and 2G
	(blue) RGB stars of the SMC cluster NGC\,121. The black box outlines
	the RGB bump region. \textit{Panel (b)}: LF of the 1G
	(red histogram) and 2G (blue histogram) stars. The peak of the
	corresponding kernel distributions marks the location of the 1G and 2G
	RGB bump. Their magnitude displacement, $\mathrm{\Delta m^{(2G,1G)}}$,
	has also been reported on the top.\label{fig1}}
\end{figure}

%\section{The RGB bump of multiple stellar populations}
%
Thanks to the selection of 1G and 2G stars and by taking advantage of the
luminosity function (LF) of the RGB stars, we were able to measure the
difference in luminosity between the RGB bump of the single stellar populations
for all the Galactic clusters and for the SMC cluster NGC\,121.  As an example,
we show in Figure~\ref{fig1} the procedure to select the RGB Bump of 1G and 2G
stars of NGC\,121. Panel (a) displays the $\mathrm{m_{F814W}}$ vs.
$\mathrm{C_{F343N,F438W,F814W} = (F343N-F438W) - (F438W-F814W)}$ pseudo
color-magnitude diagram (CMD) of the RGB of the cluster, where the 1G and 2G stars
are represented as red and blue points, respectively. The LF of the 1G and 2G
stars in the RGB bump region, outlined by the black box, has been plot in panel
(b). The peaks of the kernel functions (green and yellow curves), fitting the
distributions of the 1G and 2G stars (red and blue histogram), mark
the location of the 1G and 2G RGB bumps. The corresponding magnitude
displacement, $\mathrm{\Delta m^{(2G,1G)}} = -0.052$\,mag, has been also
reported. We repeated the same procedure to determine $\mathrm{\Delta
m^{(2G,1G)}}$ in all the other available bands for this cluster.

\begin{figure}
\centering
\includegraphics[width=.4\textwidth]{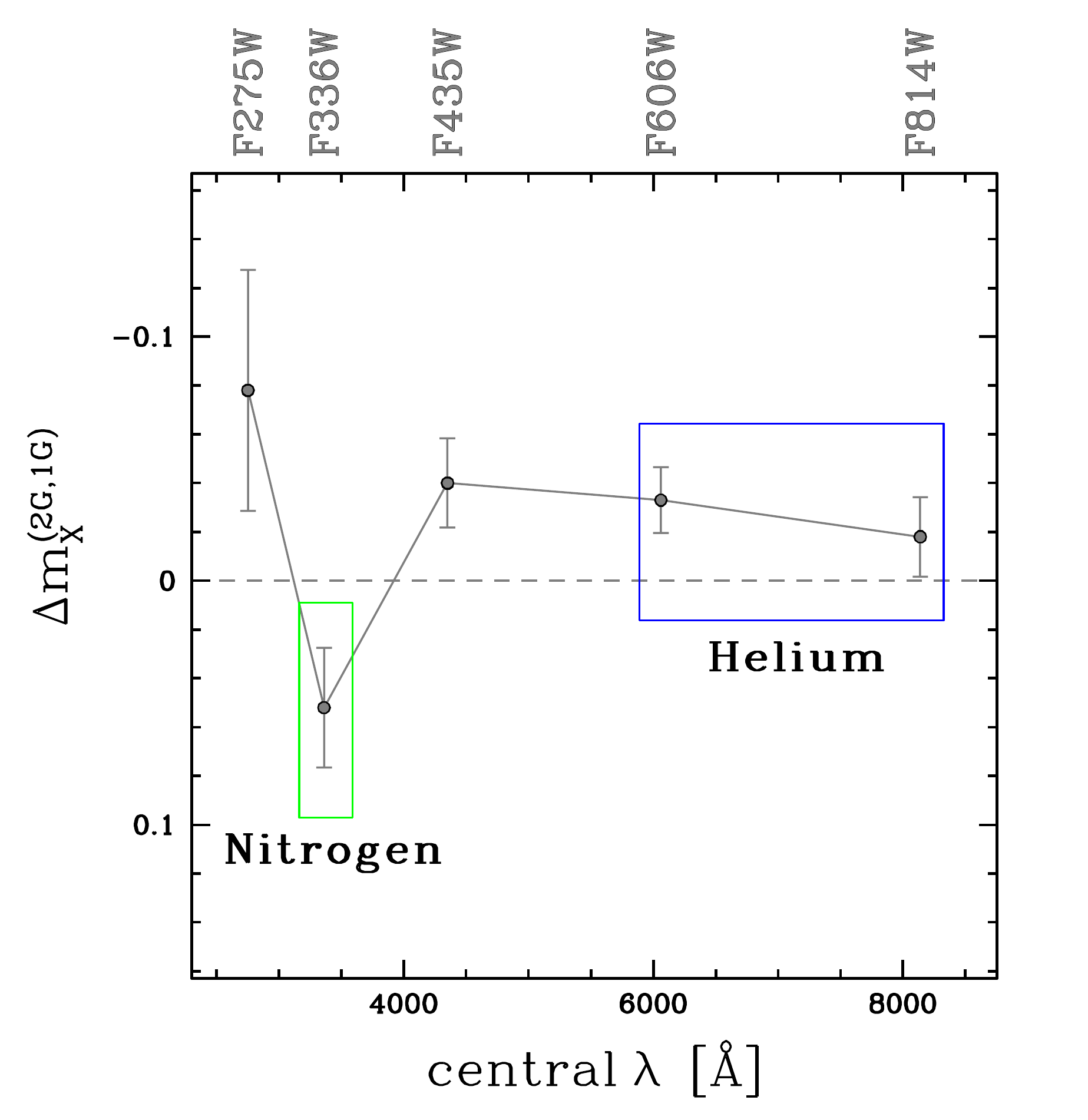}
\caption{Magnitude difference between the RGB bump of 2G and 1G stars of the
Galactic GC 47\,Tuc in five UVIS/WFC3 filters. The green and blue box
highlights the band(s) mostly sensitive to variation of nitrogen and helium,
respectively.\label{fig2}}
\end{figure}

By applying a similar procedure to all the Galactic GCs we obtained significant
measurements of the RGB bump magnitude difference for 26 out of 57 GCs.  As an
example, in Figure~\ref{fig2} we plot $\mathrm{\Delta m^{(2G,1G)}}$ as a
function of the central wavelength of the filter X, with X = F275W, F336W,
F438W, F606W and F814W, for the GC NGC\,104 (47\,Tuc). The dip visible in
correspondence of the band F336W is largely due to nitrogen enhancement of 2G
stars with respect to the 1G stars, as highlighted by the green box. On the
other hand, as a consequence of the change of the effective temperature of a
star, hence of its luminosity at a given evolutionary stage, the effects of
helium variations are mostly visible in optical bands, as highlighted by the
blue box.

\begin{figure}
\centering
\includegraphics[width=.6\textwidth]{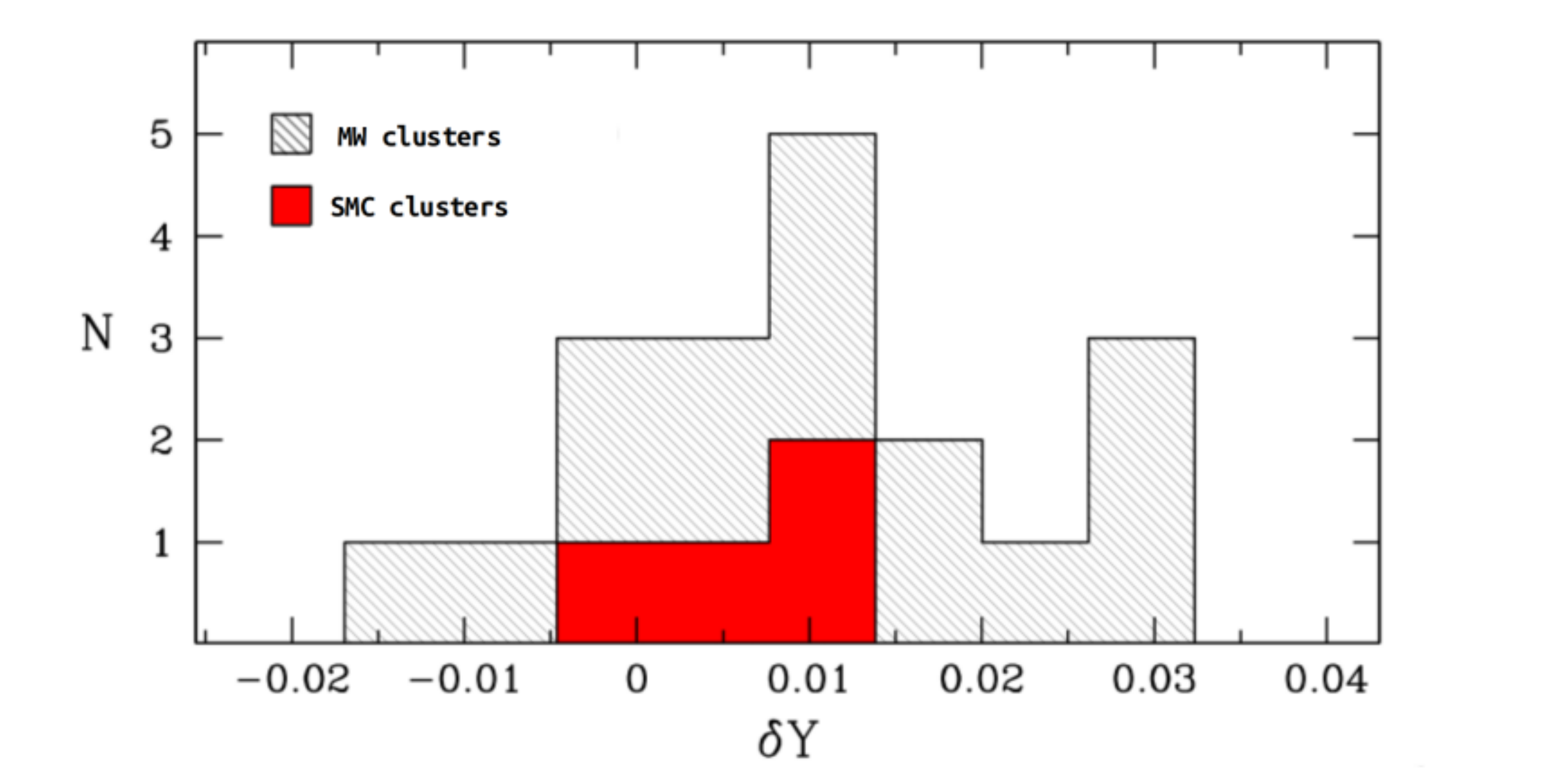}
	\caption{Distribution of the helium abundance variation between the two
	main populations of 18 Galactic GCs (\cite[Lagioia et al.
	2018]{lagioia18}; gray histogram) and of four SMC clusters 
	(\cite[Lagioia et al. 2019]{lagioia19}; red histogram). \label{fig3}}
\end{figure}

By means of synthetic spectral analysis, we estimated the contribution due to
light-elements variation to the observed scatter in optical bands. Finally, by
using appropriate theoretical models, we obtained an estimate of $\delta$Y
corresponding to the observed RGB bump magnitude displacements, for all the GCs
with [Fe/H] $< -1.0$\,dex.  As shown by the gray histogram in Figure~\ref{fig3},
we found that the mean value of the distribution of the helium variation
between 2G and 1G stars, for the final selection of 18 Galactic GCs, is
$\delta$Y $\approx 0.01$. In the case of the SMC cluster NGC\,121 we found
$\delta$Y $= 0.026 \pm 0.009$.

\section{Color spread of multiple populations in SMC clusters}
Since the RGB bump is not detectable for NGC\,339, NGC\,416 and Lindsay\,1, we
adopted a different method to measure the internal helium variation for these
SMC clusters and extended the procedure also to NGC\,121.

As shown in the four CMDs relative to the cluster NGC\,121, in Fig.~\ref{fig2},
we built the ridge line of the RGB sequence of each population, in all the
possible color combinations, and measured the difference between the color of
the points along the fiducials, at a reference magnitude, $m_{\mathrm{ref}} =
m^{MSTO}_{F814W} -2$, where $m^{MSTO}_{F814W}$ is the main sequence turn-off
magnitude in F814W band. 

\begin{figure}
\centering
\includegraphics[width=.7\textwidth]{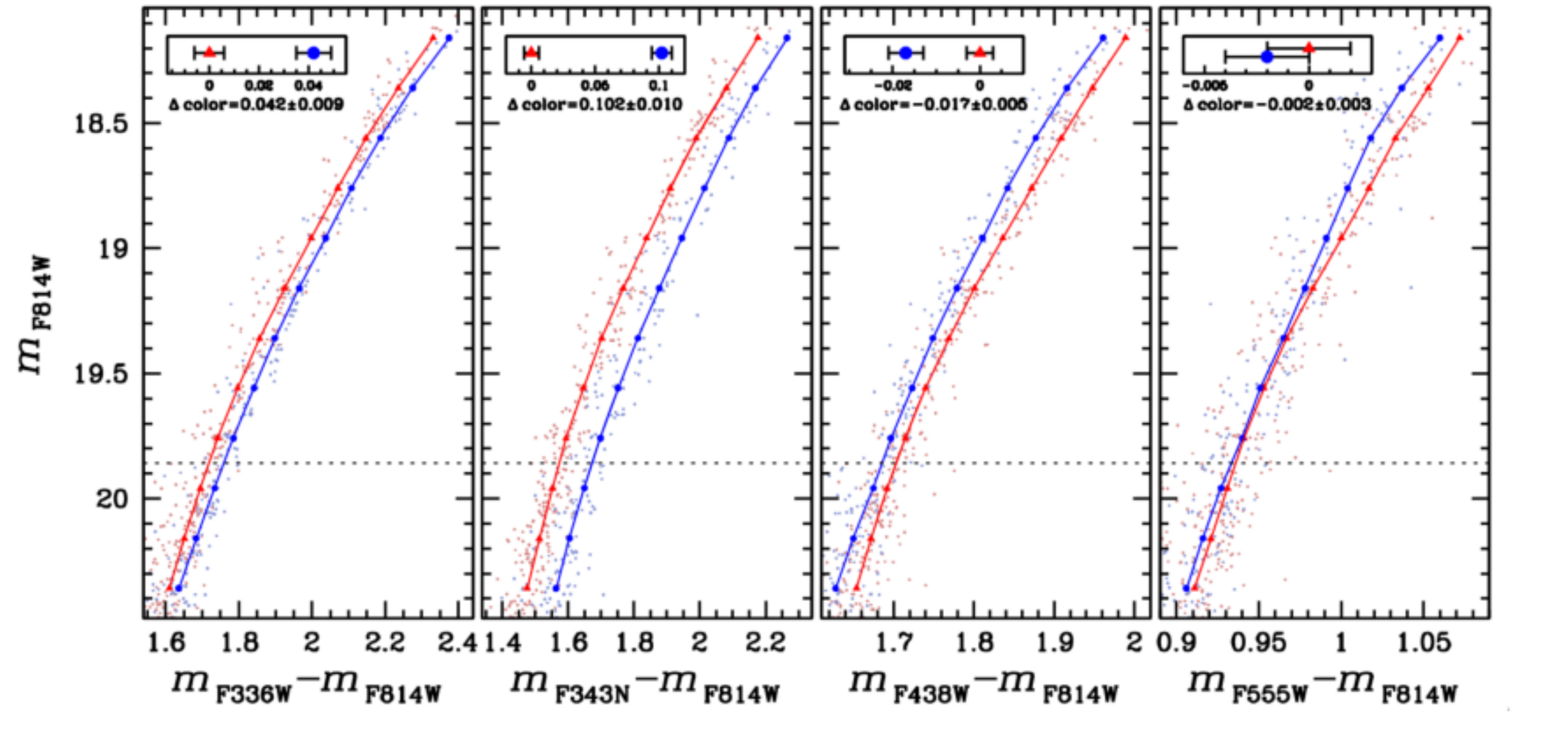}
\caption{CMDs in different color combinations, for the RGB stars of the cluster
	NGC\,121. Red and blue dots indicate, respectively, the 1G and 2G stars
	of the cluster. The corresponding fiducial lines have been overplot,
	by using the same color code. The dashed line mark the reference
	magnitude, $m_{\mathrm{ref}}$, along the fiducial lines. The inset in
	each panel displays the difference in the relative color between the
	points at $m_{\mathrm{ref}}$, along the two fiducial lines. \label{fig4}} 
\end{figure}

As done in the case of the RGB bump, we determined, by means of synthetic
spectra analysis, the contribution of light-elements variation to the observed
color spread. Finally, appropriate theoretical models allowed us to estimate
the difference in helium mass content between the two main stellar populations
corresponding to the color difference observed in optical bands. 

The red histogram in Fig.~\ref{fig3} shows that the four SMC clusters have
helium variations similar to those observe for the Galactic GCs, with 2G stars
enhanced in helium by $\delta$Y $\sim 0.01$. In particular, for NGC\,121 we
found $\delta$Y $\sim 0.009 \pm 0.006$, a value consistent within 1.5\,$\sigma$ with
that found by using the RGB bump displacement. We also found that 2G stars in
the four SMC clusters are depleted in carbon and/or oxygen (\cite[Lagioia et al.
2019]{lagioia19})

These findings provide evidence that old Galactic and intermediate and old
extragalactic GCs show consistent internal helium variations and seem to share the same
chemical properties, with helium-rich stars being N-rich and possibly C-O poor,
thus suggesting an universal mechanism at the base of the formation of multiple
populations at high redshifts.

\end{document}